\documentclass[11pt]{amsart}
\usepackage{amsfonts,amssymb,amsthm,eucal,amsmath}
\usepackage{url,geometry}

\newcommand{\C}{\mathbb{C}}
\newcommand{\E}{\mathbb{E}}

\newcommand{\N}{\mathbb{N}}

\newcommand{\R}{\mathbb{R}}

\newcommand{\tr}{\mathrm{tr}}

\newcommand{\cB}{\mathcal{B}}
\newcommand{\cD}{\mathcal{D}}

\newcommand{\cH}{\mathcal{H}}

\newcommand{\cM}{\mathcal{M}}

\newcommand{\cW}{\mathcal{W}}

\renewcommand{\[}{$$}
\renewcommand{\]}{$$}

\def\bra{\langle}
\def\ket{\rangle}
\def\dg{\dagger}
\def\eps{\varepsilon}

\begin{document}

\title{Non-additivity of R\'enyi entropy and Dvoretzky's Theorem}
\author{Guillaume Aubrun} 
\address{Institut Camille Jordan, Universit\'e Claude Bernard Lyon 1, 43 boulevard du 11 novembre 1918, 69622 Villeurbanne CEDEX, France}
\email{aubrun@math.univ-lyon1.fr}
\author{Stanis\l aw Szarek}
\address{Equipe d'Analyse Fonctionnelle, 
Institut de Math\'ematiques de Jussieu,              
Universit\'e Pierre et Marie Curie-Paris 6,          
4 place Jussieu
75252 Paris, France  {\sl and} 
Department of Mathematics, Case Western Reserve University, Cleveland, Ohio 44106, USA 
}
\email{szarek@math.jussieu.fr}
\author{Elisabeth Werner}
\address{Department of Mathematics, Case Western Reserve University, Cleveland, Ohio 44106, USA \ {\sl and} Universit\'{e} de Lille 1, UFR de Math\'{e}matique, 59655 Villeneuve d'Ascq, France
}
\email{elisabeth.werner@case.edu}
\thanks{The research of the first named author was partially supported by  the {\itshape Agence Nationale de la Recherche} grant ANR-08-BLAN-0311-03. The research of the second and third named authors was partially supported by their respective grants from the {\itshape National Science Foundation} (U.S.A.) and from the 
{\itshape U.S.-Israel Binational Science Foundation}. The second named author thanks the organizers and fellow participants (particularly F. Brandao and C. King) of the Workshop on Operator Structures in Quantum Information
 ({\itshape Fields Institute}, July 2009), which served as a catalyst for this project.
}

\begin{abstract}
The goal of this note is to show that the analysis of the minimum output $p$-R\'enyi entropy of a typical quantum channel essentially amounts to applying Milman's version of Dvoretzky's Theorem about almost Euclidean sections of high-dimensional convex bodies. This conceptually simplifies the (nonconstructive) argument by Hayden--Winter disproving the additivity conjecture for the minimal output $p$-R\'enyi entropy (for $p>1$).
\end{abstract}

\maketitle

\subsection*{1. Introduction}

Many major questions in quantum information theory can be formulated as additivity problems. These questions have received considerable attention in recent years, culminating in Hastings' work showing that the minimal output von Neumann entropy of a quantum channel is not additive. He used a random construction inspired by previous examples due to Hayden and Winter, who proved non-additivity of the minimal output $p$-R\'enyi entropy for any $p>1$. In this short note, we show that the Hayden--Winter analysis can be simplified (at least conceptually)
by appealing to Dvoretzky's theorem. Dvoretzky's theorem is a fundamental result of asymptotic geometric analysis, which studies the behaviour of geometric parameters associated with norms in $\R^n$ (or equivalently, with convex bodies) when $n$ becomes large. Such connections between quantum information theory and high-dimensional convex geometry promise to be very fruitful. 

\subsection*{2. Notation}

If $\cH$ is a Hilbert space, we will denote by $\cB(\cH)$ the space of bounded linear operators on $\cH$, and by $\cD(\cH)$ the set of {\em density matrices} on $\cH$, i.e., positive semi-definite trace one operators on $\cH$ (or {\em states} on $\cH$, or -- more properly -- states on $\cB(\cH)$). 
Most often we will have  $\cH = \C^n$ for some $n \in \N$, and we will then write $\cM_n$ 
for $\cB(\C^n)$.
\par 
For $p \geq 1$, the  $p$-{\em R\'enyi entropy} of a state $\rho$ is defined as
\[ S_p(\rho) = \frac{1}{1-p} \; \log(\tr \rho^p) .\]
(For $p=1$, this should be understood as a limit and coincides with the von Neumann entropy.) 

A linear map $\Phi : \cM_m \to \cM_d$  is called a {\em quantum channel} if it is completely positive and trace-preserving. 
The {\em minimal output $p$-R\'enyi entropy} of  $\Phi$ is then defined as
\[ S_p^{\min}(\Phi) = \min_{\rho \in \cD(\C^m)} S_p(\Phi(\rho)). \]

\subsection*{3. The Additivity Conjecture} 

The {\em Additivity Conjecture} \cite{AHW} asserted that the following equality held for every pair 
$\Phi$, $\Psi$ of  quantum channels 
\begin{equation} \label{additivityconjecture} S_p^{\min}(\Phi \otimes \Psi) \stackrel{?}{=} S_p^{\min}(\Phi) + S_p^{\min}(\Psi) .\end{equation}
 The most important case, $p=1$, has been shown to be equivalent to a number of central questions in quantum information theory \cite{shor}. Of course, had the conjecture been true for every $p>1$, it would have held also for $p=1$ by continuity.

The conjecture has been recently disproved for {\em all} values of $p \geq 1$. Early (explicit) counterexamples for $p > 4.79$ were due to Holevo and R. F. Werner \cite{HoWe}. Subsequently, the case $p>1$ was settled  by Hayden and Winter in \cite{HW}, and finally Hastings found a counterexample to the additivity conjecture for $p=1$ \cite{hastings}. The  latter two papers used nonconstructive methods; see section 9 for more comments on this aspect of the story. Hastings' presentation was rather concise, but more detailed expositions of his approach can be found, e.g.,  in \cite{FKM,BH,FK}.
\par
We want to show in this note that a large part of the analysis by Hayden and Winter is actually a fallout of Dvoretzky's theorem, a classical result in high-dimensional convex geometry dating to the 1960s \cite{dvoretzky, milman}. We note that this approach, at least in its present form, does not cover Hastings' construction.

\subsection*{4. Multiplicative form}

It will be more convenient to study a multiplicative version of the conjecture, already 
considered in \cite{AHW}. Instead of the R\'enyi entropy, we will work with the Schatten $p$-norm $\| \sigma \|_p = \big( \tr (\sigma^\dg\sigma)^{p/2} \big)^{1/p}$. (The limit case $\|\cdot \|_\infty$ is the operator, or ``spectral," norm.)  

If  $p>1$ and $\rho$ is a state, then
$S_p(\rho) = \frac{p}{1-p} \log \|\rho\|_p$, and so  the study of  $S_p^{\min}(\Phi)$ is replaced by that of 
 $\max _{\rho \in \cD(\C^m)} \|\Phi(\rho)\|_p$, or the {\em maximum output $p$-norm}.  
 The latter quantity has a nice functional-analytic interpretation:  it equals 
$\|\Phi\|_{1\to p}$, i.e., the norm of $\Phi$ as an operator from $(\cM_m, \|\cdot \|_1)$ to 
$(\cM_d, \|\cdot \|_p)$. 
This allows to rewrite conjecture \eqref{additivityconjecture} in a multiplicative form 
\begin{equation} \label{multiplicavityconjecture} 
\| \Phi \otimes \Psi \|_{1 \to p} \stackrel{?}{=} \| \Phi \|_{1 \to p} \| \Psi \|_{1 \to p} .
\end{equation}
The inequality $`` \geq"$ is trivial,  so  the conjecture asked if $``\leq"$ was always true. 

We point out that the argument that follows deals directly with the maximum output $p$-norm and 
not with  $\|\Phi\|_{1\to p}$, so the knowledge that the two are equal is not really 
needed. 
Note that it  is only obvious that the maximum output $p$-norm is equal -- for any linear map $\Phi$ -- 
to the norm  of the restriction of $\Phi$ to the $\R$-linear 
space $\cM_m^{H}$ of $m \times m$ Hermitian matrices. The fact  that 
it coincides --  for quantum channels, or even for all $2$-positive maps -- 
 with the {\em a priori} larger norm 
$\|\Phi\|_{1\to p}$ is also elementary, but less immediate \cite{Wat};  see also \cite{Aud, szarek2positive} 
for short proofs of the more general version. 

Finally, let us observe that  $\| \Phi \|_{1 \to p}$ never exceeds $1$ if  $\Phi$ is a  quantum channel. If, additionally, $\Phi $ is $\cM_d$-valued, then $\| \Phi \|_{1 \to p}$ is at least $d^{1/p-1}$ 
(the $p$-norm of the maximally mixed state in $\cM_d$).

\subsection*{5. Channels as subspaces}

Let  $\cW$ be a subspace of $\C^d \otimes \C^d$ of dimension $m$. 
Then $\Phi : \cB(\cW) \to \cM_d$ defined by $\Phi(\rho) = \tr_2(\rho)$ ({\em partial trace} in the second factor, 
i.e., $\tr_2(\rho_1\otimes \rho_2)=\tr(\rho_2) \, \rho_1$) is 
a  quantum channel.  Alternatively (and perhaps more properly), we could identify 
 $\cW$ with $\C^m$ via an isometry $V: \C^m \to \C^d \otimes \C^d$ whose range is $\cW$ and set, 
 for $\rho \in \cM_m$, 
 $\Phi(\rho) = \tr_2(V\rho V^\dg)$;  then $\Phi$ goes from $\cM_m$ to $\cM_d$.   
(One could also consider here subspaces $\cW \subset \C^d \otimes \C^r$, 
where possibly $r\neq d$; 
this would allow to preserve full generality, but would lead to more involved notation.)

By convexity, the maximum output $p$-norm, and hence also $\|\Phi\|_{1\to p}$, is attained on 
pure states. In other words
$$
\|\Phi\|_{1\to p} = \max_{x \in \C^m, |x|=1} \|\Phi(|x\ket\bra x|)\|_p,
$$
where $|\cdot |$ is the Euclidean norm.
A standard and well-known argument shows that  eigenvalues of $\Phi(|x\ket\bra x|)$ are exactly 
squares of $s_j(x)$, the ``Schmidt coefficients" of $x$, so 
$$
\|\Phi\|_{1\to p} = \max_{x \in \cW, |x|=1}\Big( \sum_{j=1}^d s_j(x)^{2p}\Big)^{1/p}
=\max_{x \in \cW, |x|=1} \|x\|_{2p}^2 \, ,
$$
where in the last expression we identify $x \in \cW \subset \C^d \otimes \C^d$ (or, to be more precise, 
$\C^d \otimes \overline{\C^d}$ --- a distinction we will ignore) with an element of 
$\cM_d$ via the canonical map induced by $u\otimes v \to |u\ket\bra v|$. (Schmidt coefficients of an element of $\C^d \otimes \overline{\C^d}$ become singular values of the corresponding element of $\cM_d$.) 
\par
In other words, $\|\Phi\|_{1\to p}$ is the square of the maximum of the ratio $\|x\|_{2p}/\|x\|_2$ over the $m$-dimensional subspace of $\cM_d$ that corresponds to $\cW$ under the canonical identification, 
and that we will still call $\cW$,
\begin{equation}\label{norm-ratio}
\|\Phi\|_{1\to p} = \max_{x \in \cW}\left(\|x\|_{2p}/\|x\|_2 \right)^2 \, .
\end{equation}  

\subsection*{6. The Hayden--Winter counterexample}

The Hayden--Winter construction can be described as follows. Let $V : \C^m \to \C^d \otimes \C^d$ be a random isometry (chosen with respect to the Haar measure) and $\Phi : \rho \mapsto \tr_2(V\rho V^\dagger)$ be the corresponding quantum channel from $\cM_m$ into $\cM_d$. 
We show in the next section that Dvoretzky's theorem implies that for $m \sim d^{1+1/p}$, such random quantum channel typically satisfies  
\begin{equation}\label{single}
\|\Phi\|_{1\to p} \sim d^{1/p-1}.
\end{equation}
Here, and throughout the remainder of the paper, $\sim$ means ``equivalent up to a universal  multiplicative constant."

Take as the second channel the (complex) conjugate channel $\bar{\Phi}$ and let $|\psi \rangle$ be the maximally entangled state in $\C^m \otimes \C^m$. It is shown in \cite{HW} (Lemma 3.3) that $(\Phi \otimes \bar{\Phi})(| \psi \rangle \langle \psi |)$ has an eigenvalue  $\geq m/d^2$, which implies that with the above choice of $m$,
$$
\| \Phi \otimes \bar{\Phi} \|_{1 \to p} \geq \| \Phi \otimes \bar{\Phi} \|_{1 \to \infty} \geq m/d^2  \sim d^{1/p-1}.
$$
On the other hand, again
with the same choice of $m$, by \eqref{single}
\[ \| \Phi \|_{1 \to p} = \| \bar{\Phi} \|_{1 \to p}  \sim d^{1/p-1} ,\]
and thus 
\begin{equation}\label{violation}
 \| \Phi \|_{1 \to p}\  \| \bar{\Phi} \|_{1 \to p} \sim \left( d^{1/p-1}\right)^2 \ll d^{1/p-1},
\end{equation}
so that we obtain a violation of the multiplicativity provided that $d^{1/p-1} \leq 1/C$, i.e., $d \geq C^{p/(p-1)}$, where $C$ is the absolute constant hidden behind the $\sim$ symbol. Moreover, this violation is asymptotically extremal. Indeed, while the inequality $\| \Phi \otimes \bar{\Phi} \|_{1 \to p} \leq \| \Phi \|_{1 \to p}=\| \bar{\Phi} \|_{1 \to p}^{-1}  \| \Phi \|_{1 \to p} \| \bar{\Phi} \|_{1 \to p}$ always holds (this follows from results from \cite{AHW}), in this example the reverse inequality also holds up to an absolute multiplicative constant. At the same time, $\| \bar{\Phi} \|_{1 \to p}^{-1}  \sim d^{1-1/p}$  is of largest possible order  in the class of $\cM_d$-valued quantum channels, see the observation at the end of section 4.

The lower estimate for $\| \Phi \otimes \bar{\Phi} \|_{1 \to p}$  was the relatively simple part of the argument from  \cite{HW}; the authors referred to the proof of their Lemma 3.3 as ``an easy calculation." 
Of course, it is ``easy" only after the fact, and the crucial point was coming up with the right 
pair of channels to analyze.

Finally, as pointed out in \cite{hastings}, the random approach allows working initially with real spaces ($\R^m, \R^d \otimes \R^d$ etc.) and producing channels $\Phi$ fitting into the Hayden--Winter scheme, whose representation in the computational basis is real. In particular, $\Phi = \bar{\Phi}$, so we have channels (acting on complex spaces) for which 
$ \| \Phi \otimes {\Phi} \|_{1 \to p} \gg \| \Phi \|_{1 \to p}^2$ and $S_p^{\min}(\Phi \otimes \Phi) < 2S_p^{\min}(\Phi)$.

\subsection*{7. Dvoretzky's Theorem}
By (\ref{norm-ratio}),  
$\|\Phi\|_{1\to p} = \max_{x \in \cW}\left(\|x\|_{2p}/\|x\|_2 \right)^2$, where $\cW \subset \cM_d$ is an $m$-dimensional subspace.
The behavior of the ratio between the Euclidean norm and some other norm on subspaces of given dimension is a quantity that has been extensively studied in geometry of Banach spaces. The most classical result in this direction is Dvoretzky's theorem: 

\medskip \noindent {\em Given $m\in \N$ and $\eps > 0$ there is $N=N(m,\eps)$ such that, for any norm on $\R^N$ (or $\C^N$) there is an $m$-dimensional subspace on which that ratio is (approximately) constant, up to a multiplicative factor $1+\eps$.} 

\medskip \noindent 

This reveals a striking geometric phenomenon: any high-dimensional convex body, no matter how peaked it may be, has sections which are close to Euclidean balls.

For specific norms this statement can be made much more precise, both in describing the dependence 
$N=N(m,\eps)$ and in identifying the constant of (approximate) proportionality of norms. The version of Dvoretzky's theorem that is relevant here is due to Milman \cite{milman}. 
(Alternative good expositions are, for example,  \cite{FLM, MS} and \cite{pisier};
the last one presents a proof based on Gaussian analysis, which allows to bypass the --- deep and not so easy to prove --- spherical isoperimetric inequality.) 

\medskip \noindent 
{\bf Dvoretzky's theorem } (Tangible version) \ {\sl Consider the $n$-dimensional Euclidean space (real or complex) endowed with the Euclidean norm $| \cdot |$ and some other norm  $\| \cdot \|$ such that, for some $b>0$, $ \|\cdot\| \leq b |\cdot|$. 
Denote $M =  \E\|X\|$, where $X$ is a random variable uniformly distributed on the unit Euclidean sphere.  
Let $\eps>0$ and let $m \leq c\eps^2 (M/b)^2 n$, where $c >0$ is an appropriate (computable) universal constant. 
Then, for most $m$-dimensional subspaces $E$ (in the sense of the invariant measure  on the corresponding Grassmannian) we have
\[ \forall x \in E, \ \ \  (1-\eps) M |x| \leq \|x\| \leq (1+\eps) M |x|.\]
}
\noindent
{\bf Remarks.}
(i) The above result is usually stated with the hypothesis 
$ a^{-1}|\nobreak\cdot\nobreak| \leq \|\cdot\| \leq b |\cdot|$  
(for some $a,b>0$). However, the parameter $a$ does not enter into the assertion;  lower bounds on $\| \cdot \|$ are related to lower bounds on $M$, needed to obtain non-trivial values of $m$ (and the function $N(m,\eps)$ mentioned earlier) in the abstract setting.\\
(ii) Standard and most elementary proofs yield the assertion only for $m \leq c\eps^2 /\log(1/\eps) (M/b)^2 n$; the dependence on $\eps$ of order $\eps^2$ was obtained in important papers \cite{Gor, Sch}. However, for our purposes it is enough to have, say, $\eps = \frac 12$, so this aspect of the story is not important. 

\subsection*{8. Dvoretzky's theorem for Schatten classes}

 In the Hayden--Winter construction, $\cW \subset \cM_d$ is a random $m$-dimensional subspace distributed according to the Haar measure on the Grassmann manifold and 
 (cf. \eqref{norm-ratio}, \eqref{single}) we want to control the ratio $\|x\|_{2p}/\|x\|_{2}$ uniformly on $\cW$, 
where $2p=: q >2$.  Thus the context in which one needs to apply Dvoretzky's theorem is the Schatten $q$-norm on the complex space $\cM_d$ for $q > 2$, in particular 
$n=d^2$, $\| \cdot \| = \| \cdot \|_q$ and $|\cdot |=\| \cdot \|_2$, the Hilbert--Schmidt norm. This has been done, e.g., in the 1977 paper \cite{FLM} (see Example 3.3 there;  \cite{FLM} focuses on real spaces, but it is noted that all proofs carry over to the complex case).
The conclusion is that if $m \sim d^{1+2/q}= d^{1+1/p}$, then the inequality 
\begin{equation} \label{dvor}
d^{1/q-1/2} \|x\|_2 \le \|x\|_q  \le Cd^{1/q-1/2} \|x\|_2
\end{equation}
holds (for some constant $C\ge1$ that does not depend on $d$ nor --- less crucially ---  on $q$)  
for all $x$ in a typical $m$-dimensional subspace of $\cM_d$.  (If we used the normalized trace to define Schatten norms, the powers of $d$ would disappear.) Combining \eqref{dvor} with \eqref{norm-ratio} yields  that when $m \sim d^{1+1/p}$, then $\| \Phi \|_{1 \to p} \sim d^{2/q-1} = d^{1/p-1}$ for a typical $\Phi$, which are exactly the values needed for the Hayden--Winter example.
\par
For completeness, let us comment on the details of the derivation of \eqref{dvor} from 
Dvoretzky's theorem. 
What we need is to find (or estimate) the quantities $b, M$ appearing in the theorem. 
Clearly, for all $x \in \cM_d$, 
\begin{equation} \label{q-estimate}
d^{1/q-1/2} \|x\|_2 \le \|x\|_q  \le  \|x\|_2 ,
\end{equation}
which yields the value of the parameter $b=1$, the lower (trivial, and not actually needed for the multiplicativity problem) estimate from (\ref{dvor}) and, {\em a fortiori}, the bound $M \geq d^{1/q-1/2}$. The upper estimate in (\ref{dvor}) will now follow once we establish that $M$ is {\em precisely} of order $d^{1/q-1/2}$. Indeed, using (the tangible version of)  Dvoretzky's  theorem with $\eps = \frac 12$ we are then led to  $m \geq c (\frac 12)^2 (M/b)^2 n \sim (d^{1/q-1/2})^2 d^2 =d^{1+2/q}$, and to an upper estimate $\frac 32 M \sim d^{1/q-1/2}$ in (\ref{dvor}) (i.e., on a ``typical" $m$-dimensional subspace).
\par
As we mentioned above, the fact that $M \sim d^{1/q-1/2}$ is  implicit in the argument from \cite{FLM}.\footnote{It is shown in \cite{FLM} that $m \sim d^{1+2/q}$ is the optimal (i.e., the largest) dimension for which (approximate) proportionality of norms does hold.  Now, if we have had $M \gg d^{1/q-1/2}$, 
Dvoretzky's theorem 
would have yielded a nearly Euclidean subspace of dimension $m \gg  d^{1+2/q}$ 
(just repeat the argument from the preceding
paragraph with $\gg$ instead of $\sim$), 
which contradicts the optimality assertion.}  However, it is instructive 
to note that it may also be obtained by many other ``standard" methods developed in 
geometric functional analysis and in random matrix theory. 
One ({\em by far} not the easiest, but most precise, at least in the appropriate asymptotic regime) 
was used by Collins--Nechita  \cite{CN}.
A simple argument to get an upper bound for $M$ 
goes as follows. Let $X$ be a random variable uniformly distributed on the Hilbert--Schmidt sphere in $\cM_d$. It is easy to check, using an elementary $\eps$-net argument, that the expectation of $\|X\|_\infty$ is bounded by $C_0d^{-1/2}$ for some absolute constant $C_0$.
Using the (pointwise) inequality $\| X \|_q \leq \| X \|_2^{2/q} \|X\|_{\infty}^{1-2/q}$ and H\"older's inequality, we get 
\[ M = \E \|X\|_q \leq \big(\E \|X\|_{\infty}\big)^{1-2/q} \leq \big(C_0d^{-1/2}\big)^{1-2/q}=C_0^{1-2/q}d^{1/q-1/2} . \]
If we are interested in good values of numerical constants, the best possible choice is $C_0=2$ --  the same ``2" as in the Wigner's semicircle law. The needed generality and precision can be extracted -- at least in the real case -- from \cite{geman}; see also \cite{HT},  \cite{ds} (Theorem 2.11) or \cite{szarekvolume} (Appendix F) for related calculations.

\subsection*{9. Derandomization}

Similarly as the approaches to the additivity conjectures by Hayden--Winter and Hastings, 
Milman's proof of Dvoretzky's theorem relies on concentration of measure via L\'evy's lemma, and so it is highly nonconstructive. Some effort has been put recently in finding explicit subspaces satisfying the conclusion of the theorem. Of course this must depend on the choice of the 
initial norm $\| \cdot \|$. The prominent example is the case of the $\ell_1$ norm on $\R^n$, which is relevant to (classical) theoretical computer science, for example to compressed sensing. In this case the dimension of the subspace is proportional to $n$. Although no explicit construction of such a subspace exists yet, recent results are promising (see \cite{indyk,GLW,IS} and references therein). We might hope to adapt such techniques to obtain constructive counterexamples to the additivity conjectures. However,  to date the best result in this direction seems to be \cite{GHP}, 
with an explicit example which works for all $p > 2$. 
An explicit counterexample to the companion problem  concerning the range $p \in [0,1)$  is exhibited in \cite{CHLMW} (it works for $p$ close to $0$). 

\subsection*{10. Shrinking under random projections and related remarks} The Hayden--Winter construction 
requires only an {\em upper} estimate on $\|x\|_q$
for all $x$ in a (random) subspace of $\cM_d$ (cf. \eqref{norm-ratio}, \eqref{violation}). 
This observation leads to counterexamples to additivity conjecture based on a phenomenon that is conceptually simpler (even if less known) than Dvoretzky's theorem. One way to express it is as follows: 
{\sl if, in the notation of Dvoretzky's theorem,  $(M/b)^2 n=:m_0 \leq m \leq n$, then the one sided estimate 
$\|x\| \leq C \sqrt{m/n} \; b \, |x|$ holds for all $x$ in a typical $m$-dimensional subspace. }
\par
While the choice $m \sim  d^{1+1/p}$ (in the construction of a random channel) results in an extremal violation of multiplicativity, the above remark shows that similar calculations  for (e.g.) all $m \ge  d^{1+1/p}$ lead to estimates of order $m/d^2$ on all the norms $\|\Phi\|_{1\to p}, \|\bar{\Phi}\|_{1\to p}$ and $\|\Phi \otimes \bar{\Phi} \|_{1\to p}$, and so a violation occurs as long as $m/d^2$ is small enough, i.e., smaller than a certain numerical constant $c>0$.
However, it should be noted that the restrictions 
$cd^2 > m \ge  d^{1+1/p}$ still imply that $d \to \infty$ as $p\to 1$.
\par
It may be more geometrically compelling to express the phenomenon referred to above in its dual form. First, the dual reformulation of Dvoretzky's theorem states that if $K$ is a symmetric body in the 
$n$-dimensional Euclidean space, then there is (relatively large) $m_0$ such that a typical orthogonal $m_0$-dimensional projection of $K$ is approximately a Euclidean ball. 
(Determining  the threshold $m_0$ involves considering the norm, for which $K$ is the unit ball, 
and then calculating parameters $M$ and $b$ for the {\em dual} norm.)
The relaxed version states that if $m_0\leq m \leq n$, then the diameter of a typical 
$m$-dimensional projection of $K$ does not exceed $C \sqrt{m/n}$ times the diameter of $K$. 
References for these remarks are, e.g.,  \cite{milmanM*} 
 and \cite{milmanVisions} (section 2.3.1), but the phenomenon can in fact be traced back (at least) to 
\cite{Gor} or \cite{JL}, among others.


\begin{thebibliography}{10}

\bibitem{AHW}
G. G. Amosov, A. S. Holevo and R. F. Werner, 
{\em On some additivity problems in quantum information 
theory}. 
Probl. Inform. Transm. 36 (4) (2000), 25--34; 
arXiv:math-ph/0003002 

\bibitem{Aud} K. M. R. Audenaert,
{\em A note on the $p\to q$ norms of $2$-positive maps.}
Linear Algebra Appl.   430 (2009), no.  4, 1436--1440; 
arXiv:math-ph/0505085  

\bibitem{BH}   F. G. S. L. Brandao and M. Horodecki, 
 {\em  On Hastings' counterexamples to the minimum output 
 entropy additivity conjecture,}
arXiv:0907.3210

\bibitem{CN} B. Collins and I. Nechita, 
{\em Random Quantum Channels II: Entanglement of random
subspaces, R\'enyi entropy estimates and additivity problems,}
arxiv:0906.1877

\bibitem{CHLMW} 
T. Cubitt, A. W. Harrow, D. Leung, A. Montanaro and A. Winter, 
{\em Counterexamples to additivity of minimum output 
$p$-R\'enyi entropy for $p$ close to $0$.}
Comm. Math. Phys. 284(1) (2008), 281--290. 

\bibitem{ds}  K. R. Davidson and S. J. Szarek,
{\em Local operator theory, random matrices and Banach spaces.  }
In {\em Handbook of the geometry of Banach spaces,}
edited by W. B. Johnson and J. Lindenstrauss
(North-Holland, Amsterdam, 2001), Vol. 1, pp. 317--366;
(North-Holland, Amsterdam, 2003), Vol. 2, pp. 1819--1820.

\bibitem{dvoretzky}  A. Dvoretzky, 
{\em Some Results on Convex Bodies and Banach Spaces.} 
In {\em Proc. Internat. Sympos. Linear Spaces (Jerusalem, 1960)}, 
Jerusalem Academic Press, Jerusalem; Pergamon, Oxford 1961, pp. 123--160.

\bibitem{FLM}   T. Figiel, J. Lindenstrauss and V. D.  Milman, 
 {\em  The dimension of almost spherical sections of convex bodies.}
Acta Math.  139  (1977), no. 1-2, 53--94.

\bibitem{FK}   M. Fukuda and C. King, 
 {\em Entanglement of random subspaces via the Hastings bound,}
 arXiv:0907.5446
 
  \bibitem{FKM}   M. Fukuda, C. King and D. Moser, 
  {\em  Comments on Hastings' Additivity Counterexamples,}
  arXiv:0905.3697

\bibitem{geman} S. Geman,
{\it A limit theorem for the norm of random matrices}. 
Ann.\ Probab.\  8 (1980), 252--261.

\bibitem{Gor}   Y. Gordon, 
{\em Some inequalities for Gaussian processes and applications.}
 Israel J. Math.  50 (1985),  265--289.

\bibitem{GHP}   A. Grudka,  M. Horodecki and {\L}ukasz Pankowski,  
{\em Constructive counterexamples to additivity of minimum output R\'enyi 
entropy of quantum channels for all $p>2$}, 
arXiv:0911.2515

\bibitem{GLW}  V. Guruswami, J. R. Lee and A. Wigderson, 
{\em Euclidean sections of $\ell_1^N$ with sublinear randomness and error-correction over the reals.}
In {\em Proceedings of the 12th International Wrokshop on Randomization
and Combinatorial Optimization: Algorithms and Techniques (RANDOM)}, 
Lecture Notes in Comput. Sci., vol. 5171, Springer-Verlag, Berlin, Heidelberg, 2008,  pp. 444--454.
 
\bibitem{HT}
 U. Haagerup and S. Thorbj\o rnsen, 
 {\em Random matrices with complex Gaussian entries.}
  Expositiones Math. 21 (2003), 293--337. 

\bibitem{hastings}
M. B. Hastings, 
{\em Superadditivity of communication capacity using entangled inputs.} 
Nature Physics {\bf 5}, 255 (2009)
 
\bibitem{HW}  P. Hayden and A. Winter, 
{\em Counterexamples to the maximal $p$-norm
multiplicativity conjecture for all $p > 1$.}
Comm. Math. Phys. 284(1):263-280, 2008;  arXiv:0807.4753
 
\bibitem{indyk}
P. Indyk, {\em Uncertainty Principles, Extractors, and Explicit Embeddings of L2 into L1}. 
In {\em STOC'07--Proceedings of the 39th Annual ACM Symposium on Theory of Computing}, 
ACM, New York, 2007, pp. 615--620. 

\bibitem{IS}
P. Indyk and S. J. Szarek,
{\em A simple construction of almost-Euclidean subspaces of $\ell_1^N$ 
via tensor products,}  arXiv:1001.0041

\bibitem{JL}
W. Johnson and J. Lindenstrauss, 
{\em Extensions of Lipschitz mappings into a Hilbert space.}  
In {\em Conference in modern analysis and probability (New Haven, Conn., 1982),}   
Contemp. Math., 26, Amer. Math. Soc., Providence, RI, 1984, pp. 189--206.

\bibitem{milman} V. Milman, 
{\em A new proof of the theorem
of A. Dvoretzky on sections of convex bodies.} 
Funct. Anal. Appl.  {5} (1971), 28--37 (English translation)

\bibitem{milmanM*}  V. Milman, 
{\em A note on a low $M\sp*$-estimate.} 
In {\em Geometry of Banach Spaces (Strobl, 1989),}
London Math. Soc. Lecture Note Ser.,  158,
Cambridge Univ. Press, Cambridge  1990, 219--229

\bibitem{milmanVisions}  V. Milman, 
{\em Topics in asymptotic geometric analysis.} 
 In {\sl Visions in mathematics. Towards 2000. (Tel Aviv, 1999).}  
{\em Geom. Funct. Anal.},  Special Volume, Part II, 
Birkh\"auser, Basel,  2000, pp. 792--815. 
 
\bibitem{MS} V. D. Milman and G.  Schechtman, 
{\em Asymptotic theory of finite-dimensional normed
spaces. With an appendix by M. Gromov.} 
Lecture Notes in Math. 1200, Springer-Verlag, Berlin 1986.

\bibitem{pisier} G. Pisier,
{\em The volume of convex bodies and Banach space geometry.}
Cambridge Tracts in Mathematics,  94.
Cambridge University Press, Cambridge, 1989.

\bibitem{Sch}
G. Schechtman. 
\newblock {\em A remark concerning the dependence on $\eps$ in Dvoretzky's theorem.}
\newblock In {\em Geometric aspects of functional analysis (1987--88)},
\newblock  Lecture Notes in Math. 1376, Springer, Berlin, 1989, pp. 274--277. 

\bibitem{shor} P. W. Shor, 
{\em Equivalence of additivity questions in quantum information theory}.  
Comm. Math. Phys.  246  (2004),  no. 3, 453--472.

\bibitem{szarekvolume} S. J. Szarek, 
{\em The volume of separable states is super-doubly-exponentially 
small in the number of qubits}. 
Phys. Rev. A 72, 032304 (2005).

\bibitem{szarek2positive} S. J. Szarek, 
{\em On norms of completely positive maps}. 
In {\em Proceedings of IWOTA 2008,}  
Oper. Theory Adv. Appl. 202, Birkh\"auser, Basel 2009, pp.535--538; arxiv:quant-ph/0603110

\bibitem{Wat} J. Watrous, 
{\em Notes on super-operator norms induced by Schatten norms.}
 Quant. Inform. Comput. 5 (2005),  58--68.
 
 \bibitem{HoWe}
R. F. Werner and A. S. Holevo, 
{\em Counterexample to an additivity conjecture for output purity of quantum 
channels.}   
J. Math. Phys. 43(9):4353-4357 (2002); arXiv:quant-ph/0203003. 

\end{thebibliography}
\end{document}